\documentclass[aps,prd,preprintnumbers,%superscriptaddress,
floatfix,nofootinbib, twocolumn]{revtex4}

\usepackage{graphicx,amssymb,amstext,mathrsfs,dsfont,amsfonts}
\usepackage{wrapfig,boxedminipage,setspace,subfigure,epsfig}
\usepackage{amsxtra,amsmath}
\usepackage{color,latexsym,url}
\usepackage[colorlinks, %pagebackref,
            hyperindex]{hyperref} 
    \hypersetup
    {
        colorlinks,%
        citecolor=red,%
        linkcolor=blue,%
        urlcolor=black,%
    }

       \def\b  {\beta}         \def\g  {\gamma}
       \def\d  {\delta}        \def\D  {\Delta}
        
\def\l  {\lambda}

\newcommand{\calo}{\mbox{${\cal O}$}}

\def\IR{{\hbox{{\rm I}\kern-.2em\hbox{\rm R}}}}
\def\IB{{\hbox{{\rm I}\kern-.2em\hbox{\rm B}}}}
\def\IN{{\hbox{{\rm I}\kern-.2em\hbox{\rm N}}}}
\def\IC{\,\,{\hbox{{\rm I}\kern-.59em\hbox{\bf C}}}}
\def\IZ{{\hbox{{\rm Z}\kern-.4em\hbox{\rm Z}}}}
\def\IP{{\hbox{{\rm I}\kern-.2em\hbox{\rm P}}}}
\def\IH{{\hbox{{\rm I}\kern-.4em\hbox{\rm H}}}}
\def\ID{{\hbox{{\rm I}\kern-.2em\hbox{\rm D}}}}

\def\be{\begin{equation}}
\def\ee{\end{equation}}
\def\ba{\begin{eqnarray}}
\def\ea{\end{eqnarray}}

\def\half{\frac{1}{2}}

\def\ra{\rightarrow}

\newcommand{\ud}{\mbox{${\mathrm{d}}$}}
\newcommand{\dd}{\mbox{${\mathrm{d}}$}}
\def\det{{\rm det}}

\def\ea{{\it et al}. }

\newcommand{\beq}{\begin{equation}}
\newcommand{\eeq}{\end{equation}}
\newcommand{\bea}{\begin{eqnarray}}
\newcommand{\eea}{\end{eqnarray}}

\newcommand{\trho}{{{\widetilde \rho}}}

\begin{document}

\newcommand\sect[1]{\emph{#1}---}

\preprint{
\begin{minipage}[t]{3in}
\begin{flushright} SHEP-12-10
\\[30pt]
\hphantom{.}
\end{flushright}
\end{minipage}
}

\title{Holography of the Conformal Window}

\author{Raul Alvares}
\email{raulalvares@gmail.com}
\affiliation{ School of Physics and Astronomy, University of
Southampton, Southampton, SO17 1BJ, UK   \vspace{-0.2cm}
}
\author{Nick Evans}
\email{evans@soton.ac.uk}
\affiliation{ School of Physics and Astronomy, University of
Southampton, Southampton, SO17 1BJ, UK  \vspace{-0.2cm}
}
\author{Keun-Young Kim}
\email{K.Y.Kim@uva.nl}
\affiliation{ Institute for Theoretical Physics, University of Amsterdam, Science Park 904, \\ Postbus 94485, 1090 GL Amsterdam, The Netherlands
\vspace{0.2cm}}

\begin{abstract}
\noindent 
Inspired by the model of Jarvinen and Kiritsis, we present a simple holographic model 
for the on set of chiral symmetry breaking at the edge of the conformal window
in QCD in the Veneziano limit. Our most naive model enforces the QCD two loop running coupling
on a D3/D7 holographic brane system. The mass of the holographic field, describing the chiral 
condensate in the model, is driven below the BF bound when the running is sufficiently strong,
triggering chiral symmetry breaking for $N_f/N_c<2.9$. This model though contains too great a
remnant of supersymmetry and does not correctly encode the perturbative anomalous dimensions of QCD.
In a second model we impose the QCD anomalous dimension result and find chiral symmetry breaking
sets in for $N_f/N_c=4$ at a BKT-type phase transition. In this case the transition is triggered
when the anomalous dimension of the mass operator $\gamma_m=1$. 
             
\end{abstract}
\date{\today}

\maketitle

%\tableofcontents

\section{Introduction}

There has been much interest in how the phase of QCD depends on the number
of quark flavours for many years now. In the Veneziano limit, where the number of
colours $N_c \rightarrow \infty$ with fixed $x=N_f/N_c$, we may treat $x$ as a continuous variable.
At $x=11/2$ the one loop beta function vanishes. Just below that value of $x$ the theory is known to
be asymptotically free and to have
a Banks-Zak fixed point \cite{Caswell:1974gg,Banks:1981nn} at which the one and two loop beta functions balance to give a non-trivial, perturbative,
conformal, IR fixed point. As $x$ falls the fixed point value rises until the perturbative regime is lost.
Based on the observation of chiral symmetry breaking in $N_c=3, N_f=3$ QCD it is presumed that at some 
critical $x_c$ the IR  conformal theory is replaced by one with a chiral condensate and a mass gap.

A number of methods have been used to estimate $x_c$. Truncated Schwinger-Dyson equations suggest
$3.5<x_c<4$ \cite{Appelquist:1996dq,Appelquist:1998rb}. In these models chiral symmetry breaking is triggered when the anomalous dimension of the
quark anti-quark operator hits of order one\footnote{ In the Schwinger Dyson
analysis the criticality condition for chiral symmetry breaking is when $\gamma_m (2 - \gamma_m)=1$ and hence
$\gamma_m=1$. To predict the critical value $x_c$ one needs to make an assumption for the dependence of $\gamma_m$ 
on $x$ at strong coupling. In \cite{Appelquist:1998rb} the one loop form $\gamma_m^{(1)}$ is used and the criticality
condition expanded at small $\gamma_m^{(1)}$ giving $\gamma_m^{(1)}=1/2$ as the condition if extended to strong coupling. 
Using the two loop form for the running
coupling in $\gamma_m^{(1)}$ gives the prediction $x_c=4$.} \cite{Appelquist:1988yc}. The precise value for $x_c$ then depends on the truncation scheme,
the choice made
for the running coupling profile with energy scale, $\mu$, and the anomalous dimension relation for 
the quark mass term, $\gamma_m$.  Other attempts
to estimate the critical value have been made in \cite{Ryttov:2007cx}-\cite{Gies:2005as} and typically give a similar estimate. Recently there has been much interest in simulating such theories on the lattice too \cite{Aoki:2012ep}-\cite{Iwasaki:2003de} the simulations
are still in an early phase but already support the general picture from all these analyses.

Jarvinen and Kiritsis \cite{Jarvinen:2011qe}
have recently proposed a holographic model of the strongly coupled near conformal
regime around $x_c$ (the work in \cite{Jarvinen:2009fe,Antipin:2009dz,Alanen:2009na,Alanen:2010tg,Kutasov:2011fr,Kutasov:2012uq} is also very relevant). Their model consists of 5D supergravity with a dilaton field dual to the running coupling
and a ``tachyon'' field dual to the chiral condensate. They impose potentials for all these fields
that generate the known two loop running for the coupling and the perturbative relation for the anomalous
dimension in the UV of their description. They predict the range $3.7<x_c<4.2$ and that the transition 
at $x_c$ is a BKT type transition in which the condensate grows exponentially from the transition (as
expected - see \cite{Miransky:1996pd,Kaplan:2009kr}).
In a sense all this physics is imposed by the choice of potentials in the model but those choices are reasonable and it is encouraging that the results match other estimates.

In this paper we wish to attempt a similar construction using an alternative holographic model
of chiral symmetry breaking. The D3-D7 system \cite{Karch:2002sh}
provides the simplest and best understood holographic
description of a strongly coupled gauge theory with quark fields. At heart it consists of $SU(N_c)$ ${\cal N}=4$
super Yang-Mills theory with $N_f$ ${\cal N}=2$ quark hypermultiplets. In the quenched approximation the theory is
conformal and on the gravity side is described by probe D7 branes in $AdS_5 \times S^5$. The theory is 3+1 dimensional
at all energy scales and has a conformal UV in which the identification of the operator matching 
between the field theory and the gravity description is clean. The simplest description of chiral symmetry
breaking is found by imposing a background magnetic field on this theory \cite{Filev:2010pm} - the description is regular
throughout and the interpretation again clear cut. Interestingly the DBI action for the probe with the
magnetic field present is equivalent to the same theory with a particular choice of running gauge coupling.
This effective dilaton is not backreacted on the geometry.
It therefore seems natural to move to phenomenological models where one simply imposes some running coupling
on the theory by hand - the underlying reaction of the holographic description seems likely to correctly
capture the resulting physics of chiral symmetry breaking. Indeed in a recent paper we have looked at the phase structure of just such a model
with a running coupling with a step change between two conformal regimes \cite{Evans:2011eu}. 
The imposition of the two loop
QCD running is a very similar analysis which we explore here, 
concentrating though on the transition to chiral symmetry breaking. Placing the probe brane in the 
presence of the dilaton matching the two loop gauge running
essentially looks at the dynamics of one of the $N_f$ quark flavours in a background backreacted to the full
dynamics of the $N_f$ quarks. Our model is more direct than that of Jarvinen and Kiritsis \cite{Jarvinen:2011qe}
in that we simply impose 
the running of the gauge coupling, and also in a later analysis the QCD anomalous dimension relation, rather
than imposing a potential and then solving for these quantities. If one had the correct gravity dual of the
gauge theory then the more involved process would capture more of the dynamics but if we are simply modeling
the gauge theory then our approach may be sufficient.   

First we will very naively impose the two loop QCD beta-function on the D3/D7 dynamics. We find that
chiral symmetry breaking is induced for $x < 2.9$ and that the transition to the chiral symmetry breaking
phase is second order in nature. This value of $x_c$ is low relative to other estimates and the transition type
is at odds with that argued for in QCD in \cite{Miransky:1996pd,Kaplan:2009kr}. To understand this we recast the DBI action
for small fluctuations about the chirally symmetric phase as a slipping mode in AdS$_5$ (we study this 
analysis in more detail in the Appendix). One can then plot its 
mass squared as a function of the radial coordinate and seek a violation of the Breitenlohner Freedman (BF)
bound \cite{Breitenlohner:1982jf}
which would lead to an
instability. The model only shows an anomalous dimension for the chiral condensate in the regimes where
the dilaton is running and the size of the anomalous dimension is proportional to the strength of that
running. Our critical value of $x_c=2.9$ corresponds to the theory which first has sufficiently strong 
running present. In terms of the slipping mode mass squared the BF bound must be violated over a sufficient 
interval in the radial direction of the gravity description - within that interval the BF bound is substantially
violated at the transition leading to the second order behaviour.

This analysis highlights a failure of the D3/D7 system as analyzed so far - it has too much supersymmetry present.
In the IR conformal regime the background gauge dynamics returns to that of ${\cal N}=4$ Super Yang-Mills. It has too much symmetry
and does not induce an anomalous dimension for the quark mass/condensate no matter how large the gauge coupling. 
 This is in direct contradiction to QCD where the anomalous dimension $\gamma_m$ is directly proportional to the magnitude of the coupling, at least in the perturbative regime \cite{Vermaseren:1997fq}.
Simply put we need to introduce
more supersymmetry breaking into the description. We show how by a choice of background dilaton the QCD one-loop
anomalous dimension relation can be imposed on the model by hand. We next impose on top the two loop QCD running 
profile within the anomalous dimension relation. In this model the slipping mode's mass squared, $m^2$, asymptotes to
$-3$ in the UV and to some lower IR fixed point value. As it passes through the BF bound of $-4$ chiral symmetry breaking
is triggered.  
The two loop running's IR fixed point implies that at the transition the IR mass squared lies at exactly $-4$ and 
this is the condition needed for a BKT transition (see \cite{Jensen:2010ga,Evans:2010hi} for the first examples of holographic BKT transitions), which we indeed observe. In this model $x_c=4$.

 It's worth stressing that the BF bound is violated in these holographic models precisely when 
$m^2=-4$ and, using the usual conformal AdS mass-operator dimension relation, $\gamma_m=1$.
This seems a robust holographic prediction, particularly since we envisage a conformal IR regime where the AdS
mass-dimension relation is expected to hold. Note that $x_c$ and the BKT transition behaviour are completely determined
by the IR fixed point behaviour of the coupling and the precise non-perturbative running is not crucial. There are more vagueries in the precise prediction of $x_c$ since we must assume
a non-perturbative relation between the anomalous dimension and the value of the IR coupling. We have used the 
leading perturbative relation between $m^2$ and the one loop anomalous dimension and extended it to
the non-perturbative regime, giving criticality when
$\gamma^{(1)}_m=1/2$ and $x_c=4$. Given the full QCD dynamics this value may be different though.

\section{The D3/D7 System}  

Our starting point is the holographic D3/D7 system
\cite{Karch:2002sh}. 
Strings tied to the surface of the $N_c$ D3 branes generate the
adjoint representation fields of the ${\cal N}=4$ gauge theory.
Strings stretched from the D3 to the D7 are the quark fields
in the fundamental representation of the $SU(N)$ group.

In the strong coupling limit the D3 branes are replaced by the geometry that they induce. We will consider a
gauge theory with a holographic dual described by the Einstein
frame geometry AdS$_5 \times S^5$ 
\begin{equation} 
\ud s^2 = 
{r^2 \over R^2} \ud x_{4}^2 + {R^2 \over r^2} \left( \ud \rho^2 +
\rho^2 \ud \Omega_3^2 + \ud w_5^2 + \ud w_6^2 \right) \,,
\end{equation}
where we
have split the coordinates into the $x_{3+1}$ of the gauge theory,
the $\rho$ and $\Omega_3$ which will be on the D7 brane
world-volume and two directions transverse to the D7, $w_5,w_6$.
The radial coordinate, $r^2 = \rho^2 + w_5^2 + w_6^2$, corresponds
to the energy scale of the gauge theory. 
The radius of curvature
is given by $R^4 = 4 \pi g_s N \alpha^{'2}$ with $N$ the
number of colours.  The $r \rightarrow \infty$ limit of this
theory is dual to the ${\cal N}=4$ super Yang-Mills theory where 
$g_s = g^2_{\mathrm{UV}}$ is the constant large $r$ asymptotic value of the gauge coupling.

In addition we will allow ourselves to choose the profile of the dilaton as
$r \rightarrow 0$.  Simplistically this represents the running of the gauge theory coupling, 
$e^\phi \equiv \beta$,
where the function $\beta \rightarrow 1$ as
$r \rightarrow \infty$.
For the coupling profiles we will consider later 
the UV form of $\beta$ will have weak logarithmic running present - we will
impose a UV cut off when $\beta=1$ corresponding roughly to the scale where
the holographic dual should be matched to perturbative QCD. Above that cut off 
we simply set $\beta=1$. The physics we study is all in the IR
and not affected by the precise form of this cut off though. In the final section we will use 
the dilaton function as an input to the DBI action for the quark physics to
enforce the QCD anomalous dimension relation. At this point the relation between the
dilaton, the gauge coupling and phenomenological corrections to the DBI action become less
clear but our philosophy is simply to phenomenologically enforce the correct quark physics in the DBI action.  

We will introduce a single D7 brane probe into the
geometry to represent the dynamics of one quark in the theory - by treating the D7 as a probe we are
working in a quenched approximation although we can reintroduce
some aspects of the $N_f$ quark loops through the running coupling's form. 
This system has a $U(1)$ axial symmetry on
the quarks, corresponding to rotations in the $w_5$-$w_6$ plane,
which will be broken by the formation of a quark condensate. 
The hope is that the dynamics of chiral symmetry breaking
for the quark described by the probe is generic across many gauge theories and
the results will be applicable to QCD.

We find
the D7 embedding function e.g. $w_5(\varrho), w_6=0$. The Dirac Born
Infeld (DBI) action in Einstein frame is given by 
\begin{equation} \label{a1}
\begin{split}
S_{D7}  &=  -T_7 \int \ud^8\xi e^\phi  \sqrt{- \det P[G]_{ab}}  \\
 &=  -\overline{T}_7 \int \ud^4x~ \ud \rho ~ \rho^3 \beta \sqrt{1 +
(\partial_\rho L)^2} \,,
\end{split}
\end{equation} 
where $w_5\equiv L$, $T_7 = (2 \pi)^{-7} \alpha'^{-4} g^{-2}_{\mathrm{UV}} $ and $\overline{T}_7 = 2 \pi^2 T_7$ when we
have integrated over the 3-sphere on the D7. The equation of
motion for the embedding function is therefore 
\begin{equation} \label{embed}
\partial_\rho \left[ {\beta \rho^3
\partial_\rho L \over \sqrt{1+ (\partial_\rho L)^2}}\right] - 2 L \rho^3
\sqrt{1+ (\partial_\rho L)^2} {\partial \beta \over \partial
r^2} = 0 \,. 
\end{equation}
The UV asymptotics of this equation, provided the
dilaton returns to a constant so the UV dual is the ${\cal N}=4$
super Yang-Mills theory, has solutions of the form  $
w_5 = d + c / \rho^2 + \cdots $, 
where we can
interpret $d$ as the quark mass ($m_q = d /2 \pi \alpha'$) and $c$
is proportional to the quark condensate.

The embedding equation (\ref{embed}) clearly has regular solutions
$w_5=m$ when $\b$ is independent of $r$ - the flat
embeddings of the ${\cal N}=2$ Karch-Katz theory.
Equally clearly if $\partial \beta / \partial r^2$ is none trivial
in $w_5$ then the second term in (\ref{embed}) will not vanish for
a flat embedding.

There is always a solution $w_5=0$ which corresponds to a massless
quark with zero quark condensate ($c=0$). In the pure ${\cal N}=2$ gauge theory 
with $\beta=1$ this
is the true vacuum. In the symmetry breaking geometries \cite{Babington:2003vm,Evans:2011eu} this
configuration is a local maximum of the potential.

If the coupling is
larger near the origin then the D7 brane will be repelled
from the origin ending at $\rho=0$ with $L'(0)=0$.  
The symmetry breaking of these solutions is visible directly \cite{Babington:2003vm}. The $U(1)$
symmetry corresponds to rotations of the solution in the $w_5$-$w_6$
plane. An embedding along the $\rho$ axis corresponds to a massless quark
with the symmetry unbroken.
The symmetry breaking configurations though map onto the flat case at large $\rho$
(the UV of the theory) but bend off axis breaking the symmetry in
the IR.  $L(0)$, the IR quark mass, is a good order parameter for
studying the chiral symmetry breaking that also reflects the bound state masses of
the theory.

\section{Imposing the 2-loop QCD Running}

Our first analysis is straightforward. We impose the two loop running
of the QCD gauge coupling on the dilaton profile of the D3/D7 system. That running is
determined by 
\begin{equation}
\mu { d \lambda \over d \mu} =  - b_0 \lambda^2 + b_1 \lambda^3 \,,
\end{equation}
where
\begin{equation} 
b_0 = {2 \over 3} {(11 - 2 x) \over (4 \pi)^2}, \hspace{0.5cm}    
b_1 = -{2 \over 3} {(34 - 13 x) \over (4 \pi)^4} \,.
\end{equation}
In the $b_1$, we omitted a subleasing term $\calo(N_c^{-2})$ at large $N_c$.
We simply identify the radial direction $r$ with the RG scale of the field theory
and set $\lambda=\beta$.  As is well
known these equations have logarithmic running in the UV and
an IR fixed point that grows from zero as $x$ is reduced from $x=5.5$\footnote{Note 
for reference that in the usual gap equation analysis \cite{Appelquist:1996dq,Appelquist:1998rb}
the critical coupling is given by $\lambda_c=8 \pi^2/ 3$ which is first achieved in the IR
for $x_c=4$}.

Here the UV is not strictly conformal although it approaches it at weak 
coupling asymptotically. Nevertheless it is easy to look
for chiral symmetry breaking. We continue to associate massless quarks with
D7 embeddings that approach the $\rho$ axis at large $\rho$ and seek
solutions that bend off axis with that UV boundary condition. In fact the simplest
identifier of chiral symmetry breaking is to look for solutions that begin with $L'(0)=0$
and shoot out to lie below $L=0$ in the UV.  We use the value of $L(0)$, the IR quark mass,
as the order parameter for chiral symmetry breaking.

\begin{figure}[]
\centering
  \subfigure[The model of section III where we naively impose the QCD running coupling and shows
           a second order transition at $x=2.949$. ]
  {\includegraphics[width=6cm]{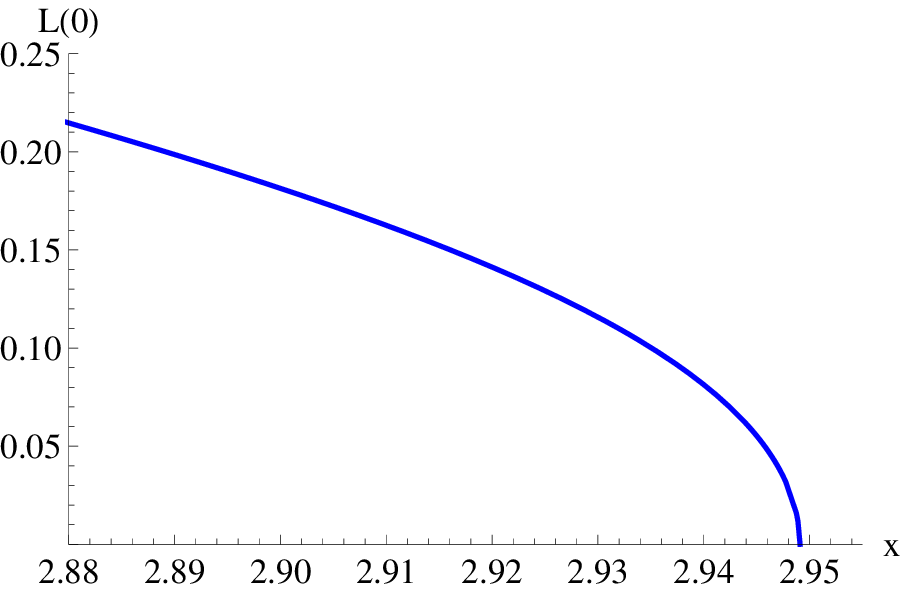}}
 \subfigure[The model of section IVa where we impose the IR
           QCD anomalous dimension relation and shows a BKT transition at $x_c=4$. ]
  {\includegraphics[width=6cm]{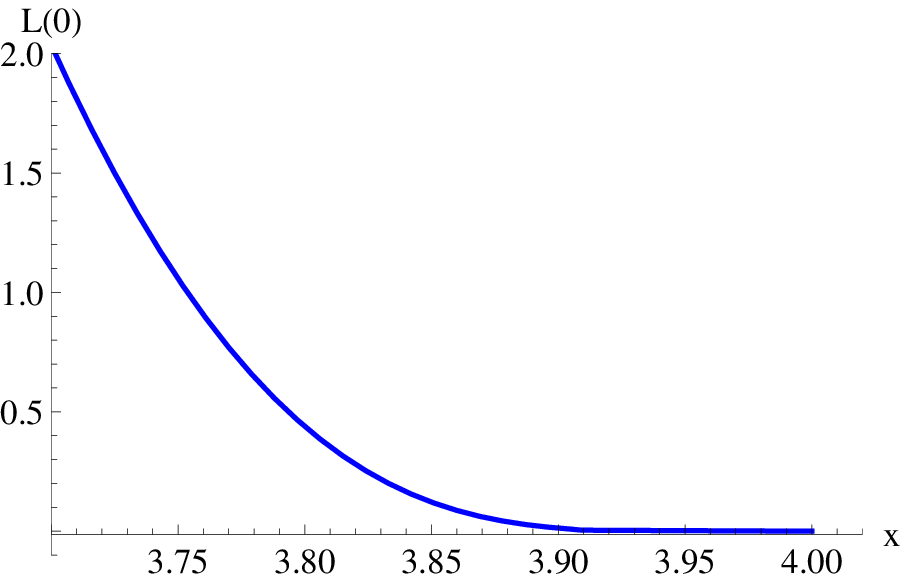}}
  \caption{
           {Plots of the IR mass L(0) against $x=N_f/N_c$ for our two models.}
           }\label{cxfirst}
\end{figure}

In Fig \ref{cxfirst}a we show a plot of $L(0)$ vs $x$. The transition is 
clearly second order and by fitting we determine it to be mean field with
critical exponent 1/2. This second order nature is of course at odds with expectations that
the transition with $x$ should be of the BKT type \cite{Miransky:1996pd,Kaplan:2009kr}. 

To understand this behaviour let us perform a linearized analysis on our DBI action
to see why the flat embedding $L=0$ becomes unstable.

We have an action, which is proportional to \eqref{a1}, 
\begin{equation} S = \int \ud \rho \lambda (r) \rho^3 \sqrt{1 + L^{'2}} \,,
\end{equation}
where $r^2 = L^2 + \rho^2$. We expand for small $L$
\begin{equation}
 S = \int \ud \rho \left( \left. {1 \over 2} \lambda (r)\right|_{L=0} \rho^3 L^{'2}   +  \rho^3 \left.{d \lambda \over d L^2}\right|_{L=0} L^2    \right) \,,
 \end{equation}
where $L' \equiv d L(\rho) / d \rho$.
To make the kinetic term canonical, we can now make a coordinate transformation\footnote {See Appendix for more detailed and general discussion on the coordinate transformation.}
 on $\rho$%
\begin{equation}
 \lambda (\rho) \rho^3 {d \over d \rho} = \tilde{\rho}^3 {d \over d \tilde{\rho}}\,, \qquad
\label{rt} \tilde{\rho} = \sqrt{{1 \over 2} { 1 \over \int_\rho^\infty {d\rho \over \lambda \rho^3}} }  \,,
\end{equation}

\begin{figure}[]
\centering
 \subfigure[The model
           with the QCD running imposed in section III ($x=3.5,3.29,3.0$). ]
  {\includegraphics[width=6cm]{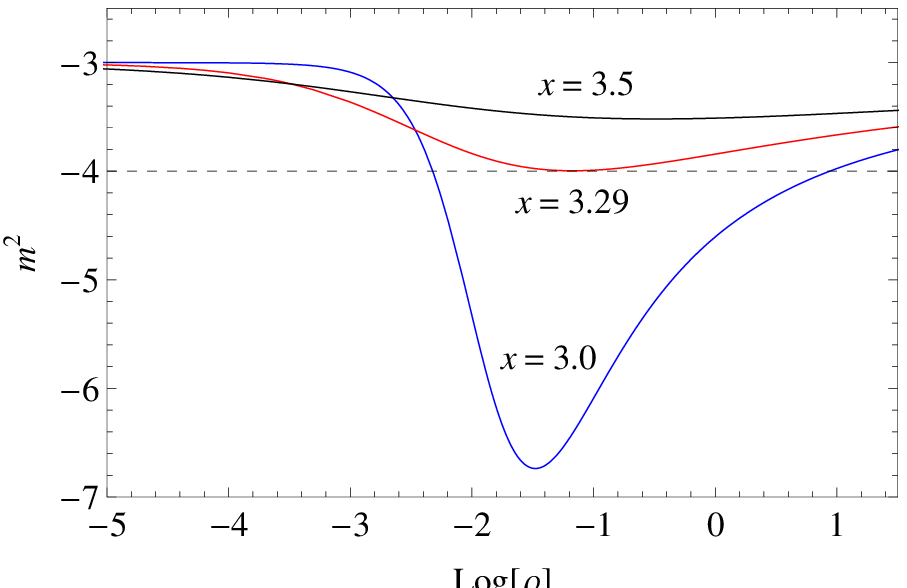}}
  \subfigure[The model
           of section IVa where the QCD anomalous dimension is imposed in the IR ($x=4.5,4,3.5,3.3$).]
  {\includegraphics[width=6cm]{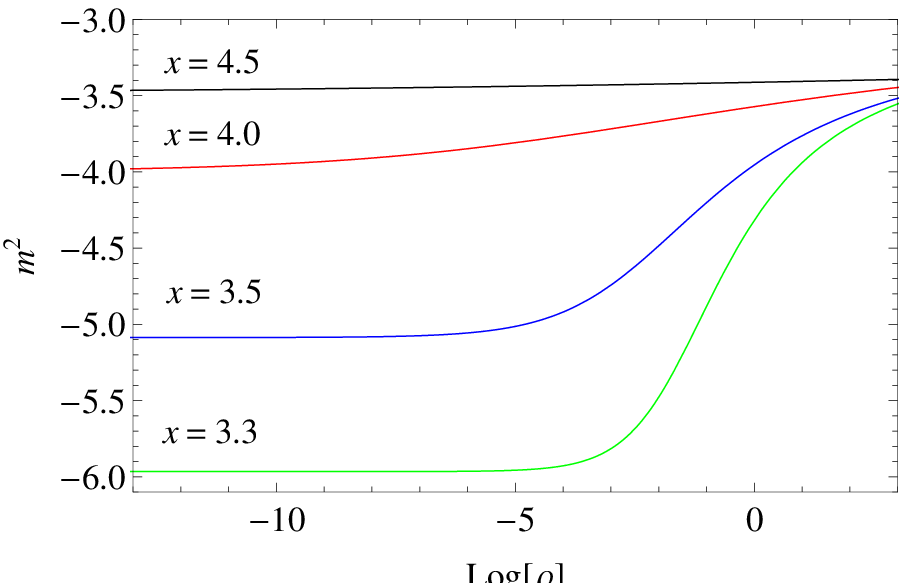}}
  \caption{
           {Plots of the AdS$_5$ slipping mode m$^2$ against r in our two models.}
           }\label{BFv}
\end{figure}

Now the first term in our action can be recast by setting $L= \tilde{\rho} \phi$
\begin{equation} \label{qa1}
 S = \int \ud \tilde{\rho}  {1 \over 2} \tilde{\rho}^3 L^{'2}  = \int \ud \tilde{\rho} {1 \over 2} \left( \tilde{\rho}^5 \phi^{'2} - 3 \tilde{\rho}^3 \phi^2 \right) \,,
\end{equation}
where $L' \equiv d L(\trho) / d \trho$.
This is the action of a canonical $m^2=-3$ scalar in AdS$_5$.
The remaining term in the action becomes
\begin{equation} \label{qa2}
S =  \int \ud \tilde{\rho} {1 \over 2} \lambda  {\rho^5 \over \tilde{\rho}}  
{d \lambda  \over d \rho}   \phi^2  \,.
\end{equation}
\medskip
So we have a AdS$_5$ scalar with $\rho$ dependent mass squared
\begin{equation} \label{deltam}
m^2 = -3  - \d m^2 \,, \qquad  \d m^2 \equiv - \lambda  {\rho^5 \over \tilde{\rho}^4  } {d \lambda  \over d \rho} \,.
\end{equation}
We plot this mass against $\rho$ in Fig \ref{BFv}a for a variety of choices of $x$. 
We first note that the mass squared
of the slipping mode approaches $-3$ in both the UV and IR, which we will return to shortly. 
The instability that is causing our phase transition with $x$ is in the intermediate period
where the mass squared is falling below the BF bound of $-4$. Note that $x=3.29$ is the first case
where the BF bound is met at one point in $\rho$ but that this value of $x$ is significantly
above the critical value $x_c=2.95$ found above. Apparently the BF bound must be violated
in a region of $\rho$ for the instability to trigger a transition. At the point of transition 
the BF bound is violated in a range of $\rho$ and the effective mass squared lies considerably
below the BF bound in the mid-region. Such a scenario has caused a second order transition. 
See \eqref{exam11} and footnote \ref{fn0} for more examples.

This plot of the mass squared of the slipping mode against $\rho$ reveals a number of failings of
our most naive model. In particular the mass squared returns to $-3$ in the IR. The reason is that the gauge 
coupling becomes constant in the IR conformal regime and the brane construction returns to that of the
${\cal N}=4$ gauge theory with quarks. The mass squared is $-3$ because the model returns to a highly
supersymmetric configuration in the IR where the anomalous dimension of $\bar{q}q$ is protected to be $3$.
This is quite unlike in non-supersymmetric QCD  where the anomalous dimension, $\gamma_m$, 
of the quark mass $m_q$
(the dimension of $\bar{q}q$ is $3 - \gamma_m$), at one loop, is
given by 
\begin{equation} \label{gamma0}
 \gamma_m^{(1)}  
= \mu {d \ln m_q\over d \mu} =  {3 \lambda \over (4 \pi)^2}  \,.
\end{equation}
At the IR fixed point one expects a non-zero $\gamma_m$. 
However, in our holographic model, using the naive scalar mass operator dimension relation, $m^2=\Delta(\Delta-4)$, we find
\begin{align} 
  \g_m &= 3- \D = 1- \sqrt{1-\d m^2} \label{gamma1} \\
  \d m^2 &= 1-(1-\g_m)^2 \label{gamma2} \,,
\end{align}
where $\d m^2$ is defined in \eqref{deltam}. Strictly
the relation $m^2=\Delta(\Delta-4)$ is valid only in conformal regimes
where the scalar mass is constant but we allow ourselves to use it slightly more liberally here.
Therefore, the model we present only conjures an anomalous
dimension in the regime in which the coupling is running ($\d m^2 \ne 0$), breaking both conformal invariance and 
supersymmetry. For a model that is precociously asymptotically free such as $x=1$ QCD this deviation of our
model from QCD is probably not so important for the phenomenology - in both cases the coupling grows rapidly
and a quark condensate is triggered. If we wish to model the transition to chiral symmetry breaking though
with changing $x$, where we leave an IR fixed point theory, it is more crucial. Our estimate of
$x_c=2.95$ is most likely an under estimate because we have not included the contribution to $\gamma_m$ 
from the absolute value of the coupling $\lambda$. 

\section{Imposing the QCD anomalous dimension} \label{sec4}

Our naive model above of the $x$ dependence of QCD suffered from an excess of supersymmetry in the
IR regime,
left over from our underlying construction. The model included the running gauge coupling but not the QCD anomalous dimension
relation. We will now enforce the perturbative QCD form of that relation (with
the two loop QCD form for $\lambda$) on the model as 
an alternative way to include the QCD physics.  That is we use the two loop relation to fix the coupling
at the IR fixed point and then use the one loop anomalous dimension relation to predict $x_c$
from the point where the slipping mode mass becomes $-4$.

\subsection{IR Physics}

Let us first consider the IR conformal regime where we want a constant non-zero 
value of $\gamma_{IR}$.   Our model predicts the slipping mode mass \eqref{deltam}
\begin{equation}
m^2_{IR} = -3 + \beta  {\rho^5 \over \tilde{\rho}^4  } {d \beta  \over d \rho}  \,,
\end{equation}
where we substitute $\beta$ for $\lambda$ since we have given up the identification of $\beta$ in the DBI action with the gauge coupling. Here we are concentrating on making $\gamma_m$ match QCD instead so the quark physics is correct.
The choice of $\beta$ that gives such a constant $m^2_{IR}$ is
$ \beta \sim {1 \over \rho^q}$ ($0 \le q < 2$)
for which we find
\begin{equation} \label{simg} 
m^2_{IR} =  -3 - \d m^2 \,, \qquad \d m^2 = {4 q \over (2 - q)^2} \,.
\end{equation}
By \eqref{gamma1}, $\d m^2$ is related to $\g_m$
\begin{equation} \label{irgam}
\gamma_m = 1 - \sqrt{1- {4 q \over (2 - q)^2}} \,.
\end{equation}
Here the use of \eqref{gamma1} is more valid than in the previous section; the scalar mass
in the action given by (\ref{qa1}) and (\ref{qa2}) is constant and by ansatz $\tilde{\rho}$ matches the
RG scale of QCD. It is possible that back-reaction between the geometry and the scalar might disturb this
relation but it seems fairly sound.
Note the conditions that $m^2_{IR}=-4$ and $\gamma_m=1$ are the same,where $q = 0.536$.

This model displays a BKT transition as $q$ is changed continuously through $\gamma_m=1$.
As usual the BKT transition occurs due to the presence of an infinite number of  
unstable, Efimov modes at the transition point \cite{Kaplan:2009kr}. We can see them here explicitly by considering the
action for static, linearized, mesonic solutions around the $L=0$ embedding. The action is
\begin{equation} S = \int \ud \rho {\rho^3 \over 2} \left( \beta L^{'2} - {\beta \over \rho^4} \dot{L}^2
+ {\partial \beta \over \partial \rho^2} L^2 \right) \,,
\end{equation}
where a prime is a $\rho$ derivative and dot a time derivative. 
If we move to the inverse $z$-coordinate ($z=1/\rho$) we have
\begin{equation} S = \int \ud z {z^{-5} \over 2} \left( \beta z^4 L^{'2} - {\beta z^4} \dot{L}^2
- z^4 {\partial \beta \over \partial z^2} L^2 \right) \,,
\end{equation}
where a prime is now a $z$ derivative. We can now write the equation of motion for a solution of the form
$L = e^{-i \omega t} z^{(1-q)/2}  \psi$ and $\beta=z^{q}$
%
%\begin{equation}
%\phi^{''} + {(q-3) \over z} \phi' +  \left( {3 \over z^2} +\omega^2 \right) \phi =0 \,,
%\end{equation}
%
%making the coordinate change $\phi = z^{(q-3)/2} \psi$ we recover 
%
\begin{equation}
-\psi^{''} + {(3 - 8q +q^2) \over 4} {1 \over z^2} \psi = \omega^2 \psi \,,
\end{equation}
which is a 1D Schrodinger equation form with a $1/z^2$ potential. This problem
is known and becomes unstable when the coefficient of the $1/z^2$ term
is equal to $-1/4$. This condition is equivalent to $\gamma_m=1$ in 
(\ref{irgam}). At that point an infinite number of unstable negative energy modes
emerge from $E=0$. At the critical value of $-1/4$ all of those modes play a role in
the transition generating the BKT transition.

This discussion so far has been restricted to the IR and a more complete model would require that
$\beta \rightarrow 1$ in the UV. A simple fix is to set $\beta = 1 + c/\rho^q$.  In this case, 
\begin{equation}
 \d m^2(\rho;q,c) = c\, q\, \rho^{-2q} (c+\rho^q) _2F_1^2\left[1,\frac{2}{q},\frac{2+q}{q}, - c \rho^{-q}\right]
\end{equation}
Its IR asymptotic behavior is 
\begin{equation}
  \d m^2 \sim {4 q \over (2 - q)^2}\left(1-\frac{1}{c(1-q)} \rho^q + \cdots\right) 
\end{equation}
which is the same as (15) with a $\rho$-dependent correction. The IR behaviour matches our discussion above.

At this point we can make a simple model to extract the critical value of $N_f$ in QCD.
The two loop QCD beta function has a fixed point at
\begin{equation} \lambda_* = {11 - 2 x \over 13 x - 34} (4 \pi)^2 \,.
\end{equation}
In the Banks-Zak regime where perturbation theory applies, 
$\gamma_{m*} = \frac{3 \lambda_*}{(4 \pi)^2}$.
the order $\lambda$ relation between 
the $\d m_*^2$ and $\g_{m*}$ is given by
\begin{equation} \label{guess}
  \d m^2_{*} \sim  2 \g_{m*}^{(1)} =  \frac{6 \lambda_*}{(4 \pi)^2}
\end{equation}
where we used \eqref{gamma2} and \eqref{gamma0} and $\g_{m*}^{(1)}$ 
denote the order $\l$ relation.

Of course we have no true idea how to continue this relation into the non-perturbative regime but following
the spirit of \cite{Appelquist:1998rb} we will simply assume (\ref{guess}) applies at all values
of the coupling. 
The holographic model tells us that the transition will occur when $m^2=-4$  $(\d m^2_* = 1)$ so
we find, using the one loop QCD anomalous dimension result
\begin{equation} 
1 = { 6 \lambda_* \over (4 \pi)^2} = 6 {11 - 2 x \over 13 x - 34}  \,.
\end{equation}
This gives $x_c = 4$.
Note that this amounts to $\gamma_{m*}^{(1)} = 1/2$, which coincides to 
the one-loop perturbative field theory computation \cite{Appelquist:1998rb}. 

 Finally we can numerically check the BKT nature of the transition as well.  We can simply set
 $\lambda=1/\rho^q$  with $q$ and $x$  related, through the IR relations (\ref{simg}) and (\ref{guess}), 
\begin{equation} \label{xq}
  \frac{4q}{(2-q)^2} = 6 \frac{11-2x}{13x-34} \,.
\end{equation} 
We then numerically solve for the D7 embedding, $L$ as a function of $\rho$. $L(0)$, the IR quark mass, is a useful order
parameter - we show the result for $L(0)$ vs $x$ in Fig \ref{cxfirst}b - the BKT type transition is apparent with $x_c=4$. Close to $x_c$ this simple model and the case $\beta=1 + 1/\rho^q$ 
coincide since the dynamics is dominated in the far IR.

\subsection{All RG scales}

To construct a full model of the RG flow in the conformal window, one should  enforce the QCD anomalous dimension formula 
(\ref{guess}) at all energy scales or $\rho$. In particular we want
\begin{equation} \label{key}
 \beta  {\rho^5 \over \tilde{\rho}^4  } {d \beta  \over d \rho} = - \frac{6 \lambda(\rho)}{(4 \pi)^2} \,.
\end{equation}

To find the associated $\beta$ one can re-arrange for $\tilde{\rho}$, 
\begin{equation}
 \tilde{\rho} = \sqrt{{1 \over 2} { 1 \over \int_\rho^\infty {d\rho \over \beta \rho^3}} }  \,,
\end{equation}
differentiate, 
and find the differential equation
\begin{equation} \label{getlambda}
{2 \over \beta \rho^3} + \partial_\rho \left[ - { 6 \lambda}  \over (4 \pi)^2 \rho^5 \beta \beta'\right]^{1/2} =0 \,. 
\end{equation}
We can solve for $\beta$ numerically by shooting from some initial value of $\rho$ and trialling various
values of the initial condition $\beta'$. Typically the true solution lies on the crossover between 
solutions that are real at all $\rho$ and those that go complex so the correct initial condition can be tuned to.
Once found the numerical solution can be tested that it is a good solution of (\ref{getlambda}) and that
it has the IR fixed point behaviour $1/\rho^q$  where $q$ and $x$ are related by (\ref{xq}).

 We can then use these solutions to solve for
$L$ as a function of $\rho$ - close to $x_c$ the results are again those in Fig \ref{cxfirst}b since the dynamics is entirely determined by the IR fixed point.

\section{Summary}

We have presented two simple holographic models of $x=N_f/N_c$ behaviour of QCD at large $N_c$. 
In our first model we imposed the QCD two loop running directly on the D3/D7 system
through a non-backreacted dilaton profile. We found chiral symmetry breaking sets in at $x_c=2.95$ at a second
order transition. The transition is expected to be at a larger value of $x$ and to be of BKT type \cite{Miransky:1996pd,Kaplan:2009kr} and we
highlighted that this discrepancy is due to the IR supersymmetry of the model forcing $\gamma_m=0$. In
a second model we imposed the perturbative QCD $\gamma_m$ relation and found a BKT transition at $x_c=4$.

Whilst these models are much less sophisticated than the very nice model
of Jarvinen and Kiritsis \cite{Jarvinen:2011qe}, in which the AdS-space backreacts 
to the running coupling and the quark condensate, we believe they highlight the key ingredients. One must 
input into the model, either directly as we do, or indirectly 
through supergravity potentials as in \cite{Jarvinen:2011qe},  
the form of the running coupling and the impact that has on the quark anomalous dimension. Since we do not have the 
true QCD dual all of this is the model builder's choice. The clear prediction from AdS is that the chiral transition
will occur when the AdS slipping mode associated to the quark condensate hits a mass squared at the IR fixed
point of $-4$, the BF bound. This corresponds to $\gamma_m=1$. The Miransky scaling or BKT nature of the transition
is then also very clear in the holographic description through the presence of Effimov modes.

\acknowledgements
RA is grateful for University of Southampton Scholarships.
NE is grateful for the support of an STFC rolling grant.  KK acknowledge support via an NWO Vici grant of K. Skenderis. This work is part of the research program of the Stichting voor Fundamenteel Onderzoek der Materie (FOM), which is financially supported by the Nederlandse Organisatie voor Wetenschappelijk Onderzoek (NWO).

\appendix

\section{Effective scalar mass and the BF mass violation}

In this appendix, we show how to identify the effective mass of the slipping mode of the probe brane in an effective AdS space in a more general context and in more detail.  
 
In general the action of the embedding $L(\rho)$, which is a function 
of only $\rho$, a holographic direction, can be written as  
\begin{equation}
  S = \int \dd \rho \, \b(r(\rho),\rho) \, \rho^{d-1} \sqrt{1+L'(\rho)^2} \,,
\end{equation}
where $r(\rho)=\sqrt{L^2+\rho^2}$ and  $d$ is an integer related to the dimension of the background and worldvolume spacetime. For example, for the D7(D5) probe brane in AdS$_5$ $\times$ $S^5$, $d=4(3)$.  We assume that 
\begin{equation}
\b(r(\rho),\rho)=
\begin{cases} 1 & \rho \ra \infty  \quad (\mathrm{UV}) \\ 
              \frac{c}{\rho^q} \,,\quad  (q \le d-1  )& \rho \ra 0  \ \ \quad (\mathrm{IR})\,,
\end{cases}       
\end{equation}
where $c$ is constant. 
The first condition comes from the fact that the slipping mode $(\phi = L/\rho)$ is a scalar in $AdS_{d+1}$ in UV. 
The second condition restricts us to an effective IR AdS space. 
When $q=d-1$, IR space is effectively $AdS_2$.

The linearized action in terms of the slipping mode reads
\begin{equation} \label{firstcase}
\begin{split}
  S &\sim  \int \dd \rho \half \b_0 \rho^{d-1} 
  \left( \rho^2 \phi'^2 + m^2 \phi^2 \right)  \\
  & m^2 = (1-d) - \frac{d \log \b_0}{d \log \rho} + 2 \frac{\rho^2}{\b_0} \left.\frac{\partial \b}{\partial L^2}\right|_{L=0} \,,
  \end{split}
\end{equation}
where $\b_0 = \b(r(\rho),\rho)|_{L=0}$.

\subsection{Effective geometry changed}

In the UV, $\b_0 = 1$ and the action \eqref{firstcase} corresponds to 
the scalar action in AdS$_{d+1}$ space with the UV mass
\begin{equation} \label{UVm}
  m^2_{UV}  = 1-d \ge -d^2/4 \,, \quad \rho \ra \infty \,.
\end{equation}
For all $d$, the BF bound is satisfied (in $AdS_{d+1}$).
In the IR, the action is written as
\begin{align} 
  S &\sim  \int \dd \rho \half \rho^{d-q-1} 
  \left( \rho^2 \phi'^2 + m^2 \phi^2 \right)  \\
  & m^2 = (1-d+q) + 2 \frac{\rho^2}{\b_0} \left.\frac{\partial \b}{\partial L^2}\right|_{L=0} \,.\label{IRm}
\end{align}
The scalar effectively lives in $AdS_{d-q}$, where $q \le d-1$.
To go further we consider two cases:  $\b = \b(r(\rho))$ and $\b = \b(r(\rho),\rho)$.\\
{\bf case 1:} For $\b = \b(r(\rho))$,
\begin{equation} \label{simple}
  \left.\frac{\partial \b}{\partial L^2}\right|_{L=0} 
 = \frac{1}{2\rho}\frac{d \b_0}{d \rho} \,, 
\end{equation}
%
%so
%
\begin{equation}
  m^2_{IR} 
   = (1-d+q) + \frac{d \log \b_0}{d \log \rho}  = 1-d\, \label{IRm1} \,.
\end{equation}
Note that the $m^2$ is the same in UV and IR. However, the stability criteria, the BF bound $-\frac{(d-q)^2}{4}$, is now changing and a function of $q$.
Therefore, if $q$ is a continuous parameter (for this purpose, let us
continue $q$ to real values), then the BKT transition occurs at 
$q=d-\sqrt{4(d-1)}$. For $d=4$, $q\sim  0.536$, which is 
the same value we obtained in section \ref{sec4}. \\
{\bf{case 2:}} For $\b = \b(r(\rho),\rho)$, 
we have to study case by case, since \eqref{simple} is not valid. As an example, let us consider D3/D7(D5) at finite $B$ and density, $\bf{d}$ \cite{Evans:2010iy,Jensen:2010vd, Jensen:2010ga,Evans:2010hi}.

\paragraph{D3/D5}
\begin{equation}
  S = \int \dd \rho \, \b(r(\rho),\rho) \, \rho^{2} \sqrt{1+L'(\rho)^2} \,,
\end{equation}
where
\begin{equation}
  \b(r(\rho), \rho) = \sqrt{1+\frac{\bf{d}^2}{\rho^4} + \frac{B^2}{r^4}}
\end{equation}
By \eqref{UVm}, the UV mass is $-2$ in $AdS_{4}$, while, by \eqref{IRm},
the IR mass is ($d=3,q=2$) 
\begin{equation} \label{aaa}
  m_{IR}^2 = - \frac{2B^2}{{\bf{d}}^2 + B^2}
\end{equation}
in $AdS_2$. The BF bound is violated at ${\bf{d}} = \sqrt{7}B$ and 
this violation by the continuous parameter, ${\bf{d}}$ or $B$, implies the BKT transition. 
If ${\bf{d}}=0$ then $m^2_{IR} = -2$, which is consistent with \eqref{IRm1}. 

\paragraph{D3/D7}
\begin{equation}
  S = \int \dd \rho \, \b(r(\rho),\rho) \, \rho^{3} \sqrt{1+L'(\rho)^2} \,,
\end{equation}
where
\begin{equation}
  \b(r(\rho), \rho) = \sqrt{1+\frac{{\bf d}^2}{\rho^6} + \frac{B^2}{r^4}} \,.
\end{equation}
By \eqref{UVm}, the UV mass is $-3$ in $AdS_{5}$, while, by \eqref{IRm}, 
the IR mass is ($d=4,q=3$) , 
\begin{equation}
  m_{IR}^2 = - \frac{B^2 \rho^2}{{\bf d}^2 } \ra 0 \,,
\end{equation}
in $AdS_2$. It satisfies the BF bound for all $B$ and ${\bf{d}}$. However, the instability is in the intermediate regime. We can see this by expanding 
the action in the regime ${\bf{d}} / B \ll \rho \ll \sqrt{B}$, of 
which linearized equation of motion is 
\begin{equation} \label{exam11}
  L'' + \frac{1}{\rho} L' + 2\frac{1}{\rho^2} L = 0 \,.
\end{equation}
The slipping mode is effectively the scalar of $m^2 = -3$ in AdS$_3$, which 
violates the BF bound. It also can be seen more directly from \eqref{firstcase}, where the second term and third term cancel out, 
leaving the first term, $(1-d) = - 3$.
Note that this instability happens only  for a large enough $B$ (or small enough $\bf{d}$) to satisfy 
the condition ${\bf{d}} / B \ll \rho \ll \sqrt{B}$. Note also that the BF mass violation is {\it finite} as we dial $B$ for a fixed $\bf{d}$, and the 
phase transition turns out
to be of mean-field type\footnote{\label{fn0}A non-mean field (but non-BKT) type transition also can be understood in the same way. In the model studied in \cite{Evans:2010np}, it can be shown that an instability can arise in the range ($ \{ (B/O)^{1/(2-\D)} ,({\bf d}/O)^{1/(2-\D)} \} \ll \rho \ll O^{1/\D} $), where $O$ is a phenomenological operator with dimension $\D$. 
This range is essentially where the operator $O$ dominates over $B,{\bf d}$.}.
(The {\it infinitesimal} violation of the BF bound as in \eqref{aaa} is a characteristic of the BKT transition.)

\subsection{Effective geometry fixed}
There is alternative way, in which we keep the UV AdS space for all $\rho$. For this, we need to redefine the coordinate system by 
\begin{equation}
  \frac{\dd \rho}{\dd \tilde{\rho}} = \frac{\b \rho^{d-1}}{\tilde{\rho}^{d-1}} \,,
\end{equation}
so 
\begin{equation}
  \tilde{\rho} = \left( \frac{1}{(d-2) \int^{\infty}_{\rho}
  \frac{\dd \rho}{\b \rho^{d-1}}}\right)^{\frac{1}{d-2}}
\end{equation}
which is defined only for $0 \le q < d-2$. Otherwise, the integral diverges and $\tilde{\rho}$ is not defined. (So, the previous examples of the D7(D3) probe in AdS$_5$ $\times$ $S^5$ cannot be analyzed in this way; $q$ is too big.) In terms of a new coordinate $\tilde{\rho}$, the action reads
\begin{equation} \label{secondcase}
\begin{split}
  S &\sim  \int \dd \tilde{\rho} \half  \tilde{\rho}^{d-1} 
  \left( \tilde{\rho}^2 \phi'^2 + m^2 \phi^2 \right)  \\
  & m^2 = (1-d) + \b_0 \b_0' \frac{\rho^{2d-3}}{\tilde{\rho}^{2d-4}} \,,
  \end{split} \,
\end{equation}
where we consider only the case $\b = \b(r(\rho))$, so that we can use \eqref{simple}. 
Note that for a function $\b_0 = \rho^{-q}$ (the normalization of $\b_0$ does not matter, since any normalization factor is canceled in \eqref{secondcase}), $m^2$ is constant:
\begin{equation}
  m^2 = (1-d) - q\left(\frac{d-2}{d-2-q}\right)^2 \,.
\end{equation}

One might wonder if this analysis is consistent with the previous one (Appendix A.1). For example, for $d=4, q=1$, both analyses are applicable. They must be consistent since the BF bound analysis is an effective tool and how to interpret the action should not change the physics. i.e. for $d=4, q=1$, we can interpret the action of either (1) a scalar in AdS$_5$ with $m^2 = -7 $ or (2) a scalar in AdS$_4$ with $m^2 = -3$. However, both cases tell us the scalar mass violate their own BF bound, so they are consistent. 
To see this more clearly, let us check the BF bound conditions, which are 
\begin{equation}
  1-d = - \frac{(d-q)^2}{4}
\end{equation}
in Appendix A.1, and
\begin{equation}
  1-d  =  q\left(\frac{d-2}{d-2-q}\right)^2 - \frac{d^2}{4} \,.
\end{equation}
in this subsection. 
These seemingly different conditions indeed give the same results: the BF bound is violated at the value of $q = q_c$,
\begin{equation}
  q_c = d - 2\sqrt{d-1} \,.
\end{equation}

Therefore we may interpret our analysis as either (1) $m^2$ does not change but the effective background is changing (2) $m^2$ is changing but the geometry does not change. 

Of course, we can do a mixture: partial change of geometry and partial change of $m^2$. How this works in general can be seen by the following simple example. The equation of the scalar field at the boundary of AdS$_{D+1}$ space ($z \ra 0$) reads
\begin{equation}
  \Phi'' + \frac{(1-D)}{z} \Phi' - m^2 \frac{\Phi}{z^2} = 0\,.
\end{equation}
By the definition $\Phi = z^{\frac{D-d}{2}}\phi$, it can be transformed to 
\begin{equation}
  \phi'' + \frac{(1-d)}{z} \phi' -\left(m^2 -\frac{d^2 - D^2}{4}  \right)\frac{\phi}{z^2} =0\,.
\end{equation}
It is formally interpreted as the scalar in AdS$_{d+1}$ space with the modified mass $m^2-(d^2 + D^2)/4 $. In both cases the BF bound is the same, $m^2 = -D^2/4$, so the physics does not change. Especially, the equation for $D=2$ (or $d=2$) corresponding to the effective AdS$_2$ is the Schrodinger equation with the $1/z^2$ potential term and the $-1/4z^2$ potential plays a role for the BKT transition as discussed in \cite{Kaplan:2009kr} and section \ref{sec4}.

\end{document}